\documentclass[preprint,prb,amsfonts,amssymb,amsmath,floatfix]{revtex4-2}  
\usepackage{bm}        
\usepackage[]{graphicx}
\usepackage{epstopdf}
\usepackage{hyperref}
\usepackage{physics}
\usepackage{gensymb}

\begin{document}

\title{Controlling helicity-dependent photocurrent in polycrystalline Sb$_2$Te$_2$Se topological insulator thin films at ambient temperature through wave-vector of and photothermal gradient due to polarized light}

\author{Samrat Roy}
\thanks{These authors have equally contributed}
\affiliation{Department of Physical Sciences, Indian Institute of Science Education and Research Kolkata, Mohanpur, Nadia 741246, West Bengal, India}

\author{Subhadip Manna}
\thanks{These authors have equally contributed}
\affiliation{Department of Physical Sciences, Indian Institute of Science Education and Research Kolkata, Mohanpur, Nadia 741246, West Bengal, India}

\author{Chiranjib Mitra}
\email[E-mail: ]{chiranjib@iiserkol.ac.in}
\affiliation{Department of Physical Sciences, Indian Institute of Science Education and Research Kolkata, Mohanpur, Nadia 741246, West Bengal, India}

\author{Bipul Pal}
\email[E-mail: ]{bipul@iiserkol.ac.in}
\affiliation{Department of Physical Sciences, Indian Institute of Science Education and Research Kolkata, Mohanpur, Nadia 741246, West Bengal, India}
\date{\today}

\begin{abstract}

Optical control of helicity-dependent photocurrent in topological insulator Sb$_2$Te$_2$Se has been studied at room temperature on dominantly c-axis oriented granular polycrystalline samples grown by pulsed laser deposition technique. Strong spin-orbit coupling and spin-momentum locking make this system unique for their applications. We observed that photocurrent can be controlled by exciting the sample with different circular and linear polarized light, yielding a polarization-dependent current density which can be fitted very well with a theoretical model. Magnitude of the photocurrent is higher even at room temperature, compared to previous reports on other single-crystal topological insulators. Comparison with the theoretical model suggests that photocurrent has different contributions. Study of dependence of photocurrent on the angle of incidence (wave-vector) of the excitation laser beam with respect to the surface normal of the sample helps to identify origins of different terms contributing to the observed photocurrent. Incidence-angle driven helicity switching, which is a very simple and effective technique to control the directional photocurrent, has also been observed in this study. This photocurrent can also be controlled with the help of photothermal gradient generated by the excitation light beam. Enhancement and inversion of this photocurrent in presence of photothermal gradient for light incident on two opposite edges of the sample occur due to selective spin state excitation with two opposite (left and right) circularly polarized light in presence of the unique spin-momentum locked surface states. These observations renders this polycrystalline material to be more important in polarization-dependent photodetection applications as well as for spin-optoelectronics under ambient conditions.

\end{abstract}

\pacs{}
\maketitle {}

\section{Introduction}

Topological insulators (TI) are new kind of emerging quantum materials~\cite{Hasan2010,Ando2013,Moore2010} which have metallic surface states showing linear dispersion, like Dirac electrons, over the insulating bulk~\cite{Xia2009,Chen2009}. Owing to the strong spin-orbit coupling present in these systems, surface-carrier spin is locked with its linear momentum~\cite{Hsieh2009,Hsieh2009a}. This kind of spin-momentum locking yields helical surface states in the momentum space, such that electrons with one  spin orientation flows in a particular direction unhindered in absence of any nonmagnetic impurity or external magnetic field. These states are termed as symmetry protected topological states, as they are protected from back-scattering from any nonmagnetic impurity due to the presence of time-reversal symmetry~\cite{Roushan2009}. At equilibrium, there is no net charge current, although there is a pure spin current. The system can be dragged out of equilibrium by means of electrical~\cite{Tang2014,Vaklinova2016,Li2014} or optical excitations~\cite{McIver2011,Qu2018,Pan2017}. The latter generates a spin-polarized charge current at the surface by creating a spin imbalance owing to the spin-selective transitions when we use circularly polarized light for excitation. It is well understood that the circularly polarized light carrying finite photon angular momentum can selectively couple and excite surface spins~\cite{Junck2013}. Eventually the helicity can be tuned by changing the polarization of light, switching from left circular polarization (LCP) to right circular polarization (RCP). Utilizing this phenomenon of polarization-dependent photocurrent, efforts are underway to implement this in the field of spintronics and optoelectronics. 

Bismuth based chalcogenides like Bi$_2$Se$_3$, BiSb$_2$Te$_3$, Bi$_2$Te$_3$ have shown helical photoresponse and have been studied extensively~\cite{McIver2011,Yan2014,Whitney2016,Besbas2016,Zhu2014}. Recently it is pointed out that nanoflakes of Sb$_2$Te$_2$Se (STS) exhibit very high photoresponse as well as large photoconductive gain in response to visible light compared to other materials~\cite{Huang2017}. However, to the best of our knowledge, there are no reports on the helical photocurrent on STS till date. Moreover, most of the other electrical and optical measurements are limited to the single crystals, nanoribbons, nanoflakes etc.~\cite{Huang2017,Huang2017a,Lee2016} which are difficult to synthesize and are also not conducive for device fabrication where generally thin films are preferred.

Here we report a less studied, c-axis oriented polycrystalline STS thin film prepared through pulsed laser deposition (PLD). These samples have been shown to mimic truly the robust helical surface states~\cite{Lin2011}. A detailed study of photocurrent response for different polarization states (circular and linear) of excitation light beam under ambient conditions allows us to differentiate between surface and bulk contributions. Measurements with variation of the angle of incidence of the excitation beam reveals the microscopic origin of different components of helical photocurrent, like circular photogalvanic effect and  linear and circular photon drag effects~\cite{Pan2017,Okada2016,Ganichev2003,Plank2016,Mikheev2018}. Observed results have matched well with previously reported theoretical proposals. After confirming the existence of surface states and its contribution to the photocurrent, we have shown that the helical photocurrent can also be tuned with the help of thermal gradient produced by the excitation beam. Enhancement and reversal of photocurrent when the excitation beam is incident on two opposite edges further improves our understanding of the contribution from the spin-momentum locked surface states. These suggest that thin films of TI materials could constitute a huge platform for making sensors and novel spintronics based devices~\cite{Zhang2016,Pesin2012,Sharma2017,Liu2017}.

\section{Experimental Methods}
\subsection{Material preparation}

The STS thin films were deposited on quartz substrates using PLD technique~\cite{Gopal2015}, which being a nonequilibrium process has the advantage of maintaining stoichiometry of the target composition in the thin films and is capable of forming samples which may not otherwise form through conventional routes including single crystal growth techniques~\cite{cmitra2001}. PLD-grown films also have an added advantage that apart from being inexpensive it offers a unique opportunity of depositing multilayers of TI with ready-integration to layers of superconducting or magnetic materials for devices applications. Before deposition, the substrates were cleaned first with acetone and then in deionized water, where it was shaken in ultrasonic bath for 15 minutes in each case. Target material consists of pure Sb ($99.999\%$), Te ($99.999\%$), and Se ($99.999\%$) in the ratio $2:2:1$. The films were deposited through ablation of the target by an UV KrF excimer laser source ($\lambda = 248$~nm). Base pressure of the deposition chamber was around  $2 \times 10^{-5}$~mbar. To achieve best-quality thin films, substrate temperature and laser fluence were optimized. It was found that substrate temperature of $250\degree$C and laser fluence at around 1.2~J\,cm$^{-2}$ were the best conditions for high-quality film growth. Whole deposition process was carried on with continuous flow of argon gas with a chamber partial pressure of 0.75~mbar to get the appropriate plume shape. Laser pulse frequency was maintained at 2~Hz as required for epitaxial growth~\cite{Gopal2015}. After deposition, films were annealed at the same temperature for one hour which ensured high crystallinity and reduced surface roughness.

\begin{figure}[htb]
	\centering
	\includegraphics[clip,height=5.5cm]{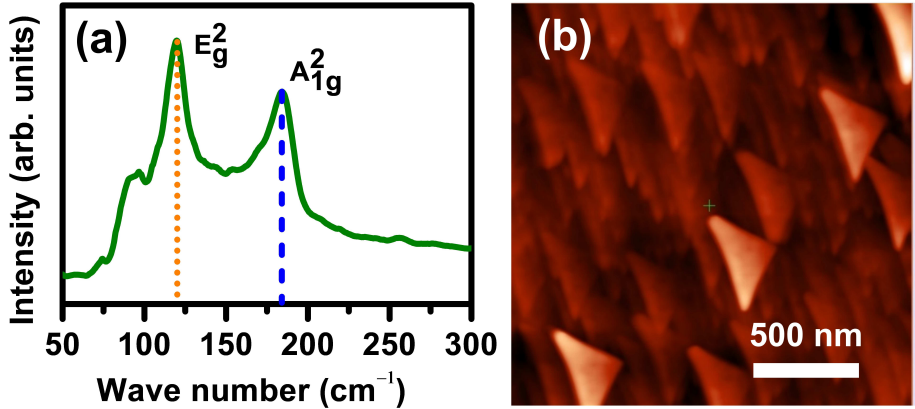}
	\caption{(a) A Raman spectrum and (b) an AFM image of the sample.}	\label{fig1}
\end{figure}

To characterize as deposited thin films, we performed Raman spectroscopy, x-ray diffraction (XRD) and atomic force microscopy (AFM). A Raman spectrum and an AFM image of the sample are shown in Fig.~\ref{fig1}. From AFM it is clear that the film growth is satisfactory with layer by layer stacking and the grains are having sizes varying from 400-500~nm. Triangular granular structure reveals the C$_{3v}$ crystal symmetry. E$^{2}_{g}$ and A$^{2}_{1g}$ peaks in Raman spectrum, respectively, at 119.4~cm$^{-1}$ and 184.3~cm$^{-1}$ signifies that the films were in single phase. E$^{2}_{g}$ and A$^{2}_{1g}$ correspond, respectively, to inter- and intra-layer atomic vibrations.

To corroborate the film quality, we analyzed the XRD spectrum which indicated highly c-axis oriented epitaxial sample growth as shown in  appendix A. The surface morphology of STS film confirms the equilibrium growth with low surface roughness. Detailed descriptions of transport measurements carried on this sample are available in appendix B. Temperature dependence of resistance and Hall measurements showed that it is metallic with p-type character, which suggests that the Fermi level is in the valence band. The bulk carrier density is about $3\times10^{18}$~cm$^{-3}$ which is consistent with previous reports~\cite{Shrestha2017}. Weak anti-localization cusps at lower magnetic field from the magnetoconductance (MC) measurement performed at low temperature certifies the existence of the TI surface states~\cite{Gopal2017}. The MC data has been fitted to the Hikami-Larkin-Nagaoka (HLN) equation~\cite{Hikami1980} for TI and we found a good match between the theory and the experimental data.   

\subsection{Optical measurements}
\begin{figure}[htb]
	\centering
	\includegraphics[clip,height=5.2cm]{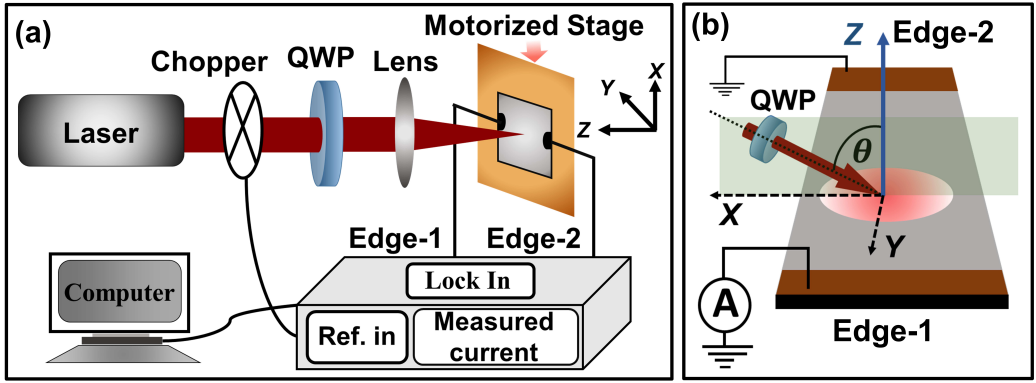}
	\caption{(a) Schematic of the polarization-dependent photocurrent measurement setup. (b) Sample geometry and transverse photocurrent measurement configuration with respect to the plane of incidence of the excitation laser beam.}	\label{schematic}
\end{figure}

The schematic of the optical experiments is shown in Fig.~\ref{schematic}. The sample was mounted on a combination of computer-controlled motorized translation and rotation stages after covering the contact on the sample by a dielectric coating. A diode laser (wavelength~$= 661$~nm) having a Gaussian beam profile was focused on the sample to a spot size of 80~$\mu$m. For the study of the  polarization-dependence of photocurrent, the laser beam was first linearly polarized. Then the polarization of the incident beam was periodically modulated by a rotating quarter wave plate (QWP) and the photocurrent was measured between the two edges, 1 and 2, fixing the laser spot at different positions on the sample along the y-direction. Photocurrent was measured by a lock-in amplifier (SR830), whose reference frequency was set at 116~Hz by a mechanical chopper that modulated the incident light intensity. With the help of rotation and translation stages, the sample could be moved in 3D in such a way that  the angle of incidence of the excitation beam on the sample could be varied from $+90\degree$ to $-90\degree$ with $0.5\degree$ precision. But the contact pads on the samples restricted us to vary angle of incidence from $+60\degree$ to $-60\degree$ only. Beam spot can traverse from edge-1 to edge-2 by translational motion of the sample with 1~$\mu$m precision.

Sample geometry of our measurements has been shown in Fig.~\ref{schematic}~(b). The excitation beam was incident in the x-z plane and the photocurrent has been measured along transverse y-direction.The angle of incidence ($\theta$) was measured with respect to the surface normal of the sample directed along z-axis. Polarization state of the excitation light beam was varied by rotating the optical axis of the QWP with respect to the polarization axis of the linearly polarized incoming beam. The rotation angle of the QWP is designated as $\Phi$.

\section{Results and Discussions}
\subsection{Controlling photocurrent by photon polarization}
\begin{figure*}[htb]
	\centering
	\includegraphics[clip,height=5.5cm]{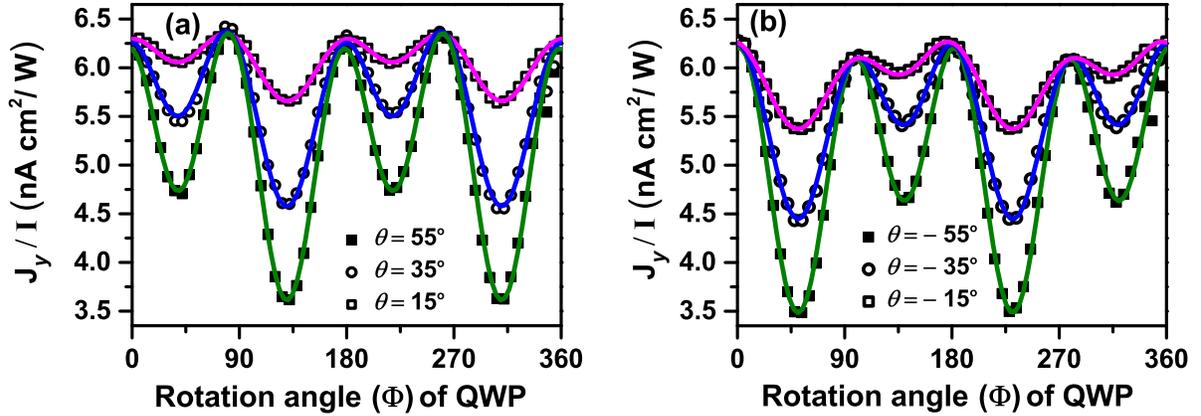}
	\caption{Photocurrent measured as a function of the rotation angle of the QWP (which changes the state of polarization of the excitation beam) for a few different angles of incidence: (a) positive angles for which light is incident from the left side of the sample normal and (b) negative angles for which light is incident from the right side of the sample normal. Data are fitted (solid lines) with Eq.~\eqref{photocurrent} described in the text.}	\label{fig2}
\end{figure*}

As described above, the prepared STS thin film was mounted on the translational stage and the photocurrent has been measured between the two edges for different positions of the laser spot on the sample. Near the edges, an enhancement of photocurrent occurs due the photothermal voltage development. Away from the edges the measured photocurrent reduces significantly due to minimization of the thermal effect. Fixing the laser spot near the center of the sample, the photocurrent measurements have been carried out while varying polarization of the incident light by rotating the QWP. Polarization-dependent photocurrent has been measured as shown in Fig.~\ref{fig2}~(a and b) for different angles of incidence $\theta$ of the excitation laser beam with respect to the surface normal of the thin film.

The variation of photocurrent with angle of rotation ($\Phi$) of the QWP (giving different states of polarization of the excitation beam) was fitted very well with the following equation~\cite{McIver2011}:
\begin{equation}
J_y = J_c\sin 2\Phi + J_1 \sin 4\Phi + J_2 \cos 4\Phi + D
\label{photocurrent}
\end{equation}
where, $\Phi$ is the angle between the optical axis of the QWP and the axis of polarization of the incident linearly polarized light. The four different components are helicity-dependent photocurrent ($J_c$), linear polarization-dependent photocurrent ($J_1$ and $J_2$) and polarization-independent thermal current ($D$). 

The spin imbalance in the surface states of the helical Dirac cone gives rise to the spin-dependent helical photocurrent. The surface states of polycrystalline TI thin films exhibit similar properties as in a single crystal which has also been confirmed from the weak antilocalization observed in transport measurements and demonstration of Shubnikov–de Haas (SDH) oscillations~\cite{Zhang2012,Zhang2012a,Gopal2017}. Theoretical reports suggested that for granular TI samples, each grain can behave as a single crystal which holds Dirac cone-like surface states~\cite{Banerjee2017}. Considering hopping from one grain to another, which can be tuned by temperature and grain size, entire film turns out to be a macroscopic TI with surface states similar to a single crystal.

\begin{figure*}[htb]
	\centering
	\includegraphics[clip,height=10.0cm]{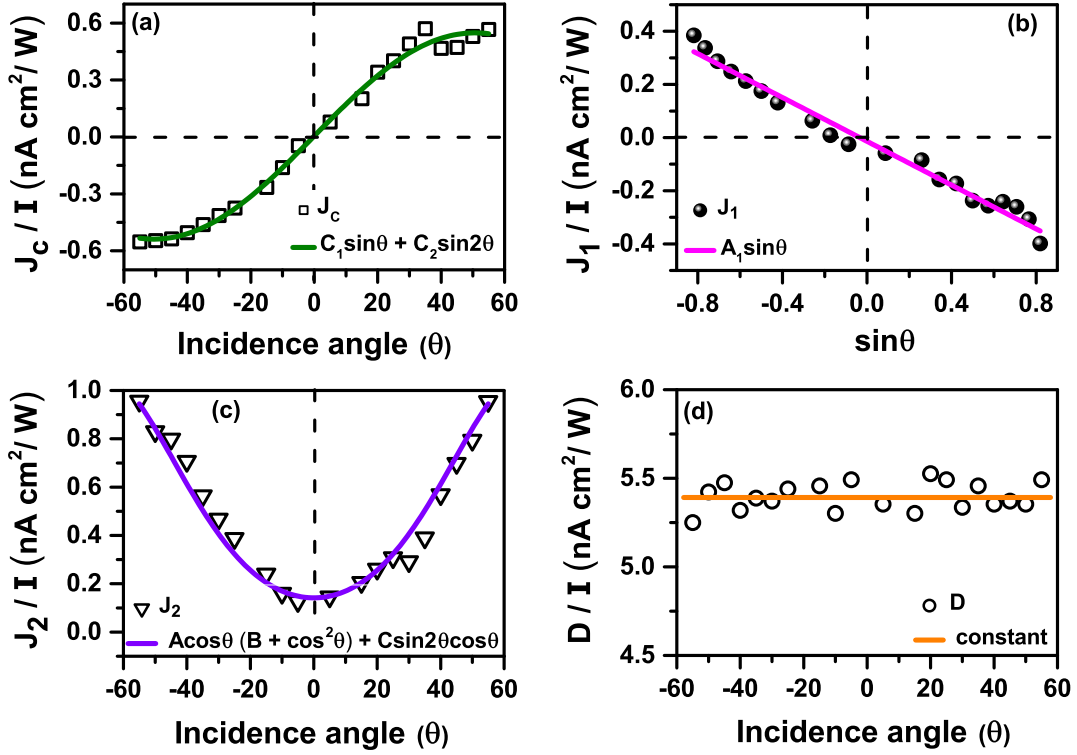}
	\caption{Variation of the strength of various components of the total photocurrent: (a) $J_c$, (b) $J_1$, (c) $J_2$, and (d) $D$, with the angle of incidence ($\theta$) of the excitation light beam. The solid lines represent fits by the functional forms written inside the respective panels.}	\label{fig3}
\end{figure*}

For better understanding of the microscopic origin of different contributions to the measured photocurrent, measurements of photocurrent have been carried out with the photon polarization varying periodically from p-polarization ($\Phi=0\degree$) to LCP ($\Phi=45\degree$) to p-polarization ($\Phi=90\degree$) to RCP ($\Phi=135\degree$) to p-polarization ($\Phi=180\degree$) for different angle of incidence ($\theta$) of the excitation beam, keeping the average laser power constant at 20~mW. The LCP and RCP states are modulated with a period of $180\degree$ and the linear p-polarization state is modulated with a period of $90\degree$. Different angle of incidence corresponds to different wave-vector of the incident light. The data are shown in Fig.~\ref{fig2}~(a and b). Strengths of various contributions to the photocurrent have been extracted by fitting the data of Fig.~\ref{fig2} with Eq.~\eqref{photocurrent}. The values of different components ($J_c$, $J_1$, $J_2$, and $D$) extracted in this way have been plotted in Fig.~\ref{fig3} as a function of the angle of incidence ($\theta$) of the excitation beam. 

Helical photocurrent arises from asymmetrically distributed spin polarization due to selective excitation in the surface states known as circular photogalvanic effect (CPGE)~\cite{Qu2018,Pan2017,Okada2016} along with linear momentum transfer to the free carriers, known as circular photon drag effect (CPDE). For semiconductor quantum wells, the CPGE emerges due to the inversion symmetry breaking at the interface. In our system, inversion symmetry has been broken at the surface, though in the bulk it remains intact. So, CPGE comes only from the surface states which comprise of Dirac electrons in TI. CPGE and CPDE can be expressed macroscopically as~\cite{Ganichev2003}
\begin{equation}
j_\alpha^{CPGE}=\sum_{\beta} \zeta_{\alpha \beta}(iE \times E^{*})_\beta
\label{eq2}
\end{equation}
\begin{equation}
j_\alpha^{CPDE}=\sum_{\beta\gamma} \eta_{\alpha \beta \gamma}q_\beta(iE \times E^{*})_\gamma
\label{eq3}
\end{equation}
where $\alpha$, $\beta$, $\gamma$ = $x$, $y$, $z$; $E$, $E^{*}$ are the electric field and its complex conjugate, $q$ is the linear momentum of the incident light and  $\zeta_{\alpha \beta},\eta_{\alpha \beta \gamma}$ are, respectively, second and third order tensors. Yu Pan \textit{et al.}~\cite{Pan2017} explained the microscopic origin of CPGE and CPDE from detailed $C_{3v}$ symmetry analysis and showed that those currents vary as $\sin\theta$ and $\sin2\theta$ respectively, where $\theta$ is the angle of incidence of the excitation light beam. Considering both these currents are present in our system, we fitted the strength of the helical component ($J_c$) of the measured photocurrent to $C_1 \sin\theta + C_2 \sin2\theta$. This fits very well with our data as shown in Fig.~\ref{fig3}~(a). Qualitatively this can be considered as having the following effect: with increasing angle of incidence, the component of photon angular momentum along the surface plane increases which enhances the coupling strength between the photon and in-plane spin angular momentum. That preferentially excites surface spins which create spin-filtered charge carriers. Due to the spin-momentum locking, this spin imbalance results in the enhancement of the helical photocurrent with increasing angle of incidence and this is also responsible for the reversal in direction of photocurrent when the beam is incident from opposite side with respect to the surface normal. 

There is not much clarity in the earlier reports about the origin of the $J_1$ component. Present work contains enough observations which claim its origin to be from the surface photogalvanic effect (SPGE)~\cite{Mikheev2018,Gurevich1993}. Anisotropic transition probability of electrons relative to the polarization direction along with the scattering from surface generates SPGE. The interband transition probability along the longitudinal direction can have the component ($E.K$)$^2$, where $E$ is the electric field lying in the incident plane and $K$ is quasimomentum of photogenerated electrons. Variation of polarization of the incident light creates an out of the incidence plane ($x$-$z$ plane in our experimental configuration) component (along $y$-axis) [see Fig.~\ref{schematic}~(b)]  which leads to another additional component 2Re[$E_x^*E_y^*K_x^*K_y^*$] apart from the ($E.K$)$^2$ described above. In the presence of this out of plane electric field component, detailed mathematical calculations suggested that the transverse SPGE current varies as $\sin\theta \sin4\Phi$~\cite{Gurevich1993}. Our experimental observations agree well with the theoretical $\sin\theta$ variation of $J_1$ with $\theta$ as shown in Fig.~\ref{fig3}~(b). 

From the polarization-dependent photocurrent measurements at different temperatures, J. W. Mclver \textit{et al.}~\cite{McIver2011} have shown that the $J_2$ and $D$ are emanating from the bulk. Siyuan Luo \textit{et al.}~\cite{Luo2017} also have reported identical observation by measuring the anisotropic polarization-dependent photocurrent by rotating the sample plane. They have claimed that $J_c$ and $J_1$ are symmetry independent, but $J_2$ and $D$ depend on bulk crystal symmetry. In our results, it has been clearly observed that the $J_c$ and $J_1$ vary as an odd function of the angle of incidence ($\theta$) of the excitation light beam whereas $J_2$ changes as an even function. This may be another possible way to claim their distinct origins. The bulk carriers can contribute in two different ways, one is the mundane dc thermal current and another comes from linear photon drag effect (LPDE)~\cite{Plank2016}. The dc thermal current should not vary with the angle of incidence, which is corroborated by Fig.~\ref{fig3}~(d). Detailed theoretical analysis shows that the variation of $J_2$ with the angle of incidence $\theta$  of the excitation light beam is captured by the following equation (see appendix C).
\begin{equation}
J_2(\theta) = A \cos\theta (B+\cos^2\theta )+C \sin^2\theta \cos\theta
\label{eq4}
\end{equation}
where $A$, $B$, $C$ are related to the Fresnel transmission coefficients for s- and p-polarized light. Figure~\ref{fig3}~(c) shows that the variation of $J_2$ with the angle of incidence $\theta$, extracted from our experimental data matches very well with that predicted by Eq.~\eqref{eq4}. The LPDE, where linear momentum is transferred from incident photons to excited charge carriers, also depends on the polarization-dependent photon absorption by the sample. 
The p-polarized light is expected to be absorbed strongly compared to other polarizations~\cite{McIver2011,Hecht2002}. The polarization independent photocurrent ($D$) has a constant background, whereas the $J_2$ component has a $\cos4\Phi$ modulation (see appendix C) with the polarization. The maxima of the  $\cos4\Phi$ ($\Phi$ is the angle of rotation of the QWP) coincides with  the p-polarization of the incident light. There is stronger absorption of the incident light at those angles of the QWP. The contribution to LPDE is ruled out for the surface carriers where spin splitting is present in the electronic structure. The coexisting photocurrents arising from Dirac Fermions ($J_C$ and $J_1$) and bulk carriers ($J_2$ and $D$), can aid in building devices based on the polarization-dependent photodetection by the topological insulating polycrystalline Sb$_2$Te$_2$Se thin films.

\subsection{Intensity dependence of photocurrent}
\begin{figure}[htb]
	\centering
	\includegraphics[clip,height=5.5cm]{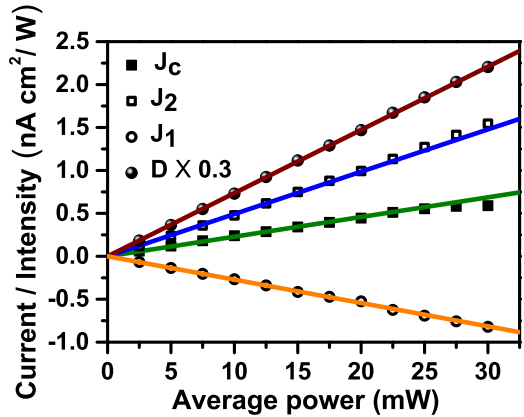}
	\caption{Variation of the components of photocurrent
(a) $J_c$, (b) $J_1$, (c) $J_2$, and (d) $D$ with the average power of the
incident laser was fitted nicely with straight lines in the 0-30~mW range. The large component $D$ is scaled down to $0.3D$ for clarity in the plot.}	\label{fig4}
\end{figure}

The above theoretical considerations were based on the assumptions that the entire photocurrent response was dependent linearly on laser intensity (i.e., second order in the electric field). Owing to high thermoelectric power~\cite{Mashhadi2016,Li2019}, the photocurrent may become nonlinearly dependent on light intensity at higher intensities. To validate that our measurements were conducted within the domain of linear dependence on light intensity, we performed intensity versus photocurrent measurements keeping the light spot at the center of the sample (to reduce the thermoelectric contribution which is dominant at the edges) and fixing the angle of incidence of the excitation beam at $50\degree$. The photocurrent measured while rotating the QWP was fitted with Eq.~\eqref{photocurrent} and we determined the variation of each of the components ($J_c$, $J_1$, $J_2$, and $D$) with intensity. It is observed that each of the components of photocurrent changes linearly in the range of 0-30~mW of average power as shown in Fig.~\ref{fig4}. Thus we have fixed our average power in all polarization-dependent measurements at 20 mW to avoid any higher order nonlinear effects and also any kind of saturation effects.

\subsection{Controlling photo-thermoelectric current by photon polarization}
\begin{figure*}[htb]
	\centering
	\includegraphics[clip,height=5.5cm]{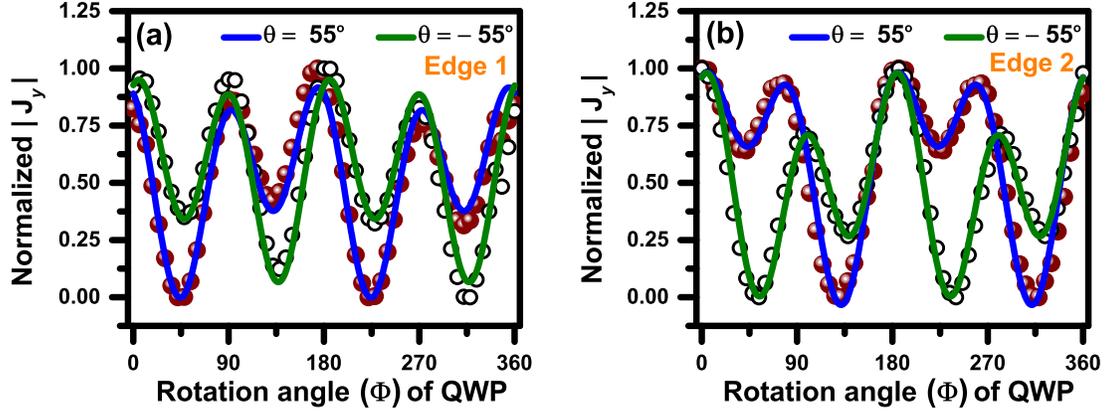}
	\caption{The magnitude of normalized total photocurrent $|J_y|$ when the laser spot was focused near edge-1 (a)  and edge-2 (b) for $\pm 55\degree$ angle of incidence. The net photocurrent flows in the opposite directions at the two edges.}	\label{fig5}
\end{figure*}

\begin{figure*}[htb]
	\centering
	\includegraphics[clip,height=6cm]{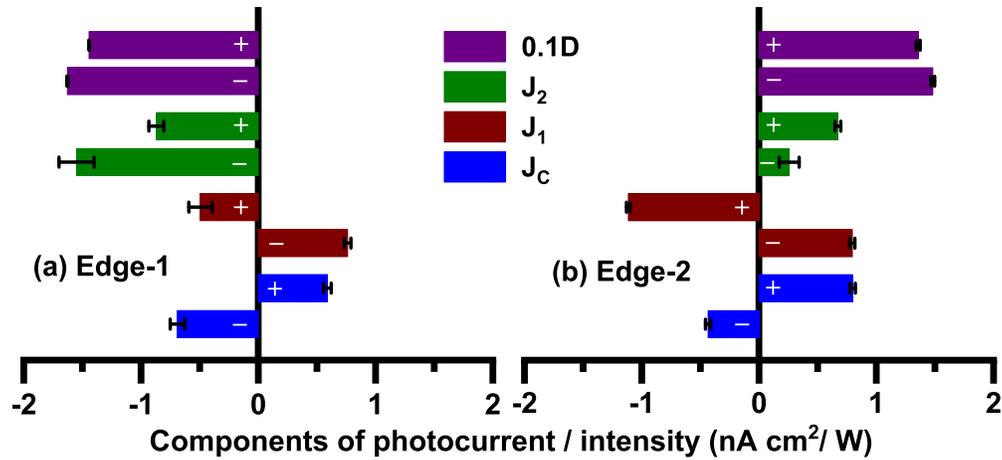}
	\caption{Comparison of various components, $J_c$, $J_1$, $J_2$, and $D$, of the measured net photocurrent $J_y$ for two opposite angles of incidence [$+$ ($-$) sign on a bar corresponds to measurement with positive (negative) angle of incidence, $\pm 55\degree$] when the laser spot was focused near the edge–1 (a) and edge–2 (b) of the sample. The large component $D$ is scaled down to $0.1D$ for clarity in the plot.}	\label{fig6}
\end{figure*}

In this section we present the study of enhancement of polarization-dependent photocurrent in presence of thermal drift for spin-momentum locked TI system. To ensure the presence of thermal gradient, we focused the excitation beam near the edges of the sample and performed the same measurement that we have described above with the incident beam at the central portion of the sample. Here, the variation of photocurrent with the rotation angle ($\Phi$) of the QWP (which changes photon polarization) has been measured as shown in Fig.~\ref{fig5}~(a and b). Fitting the measured photocurrent with Eq.~\eqref{photocurrent} gives different components of the total current for different angles of incidence. Similar variation with angle of incidence has been observed as in the case when the beam was incident on the central position except that excitation near the edge had yielded a significantly higher value of $D$. This confirms that the thermal effect originating from the bulk carriers dominates near the edges. We compared in Fig.~\ref{fig6}, the different photocurrent components when the beam was incident at the two opposite edges for two opposite angles of incidence in each case.

\begin{figure}[htb]
	\centering
	\includegraphics[clip,height=7.0cm]{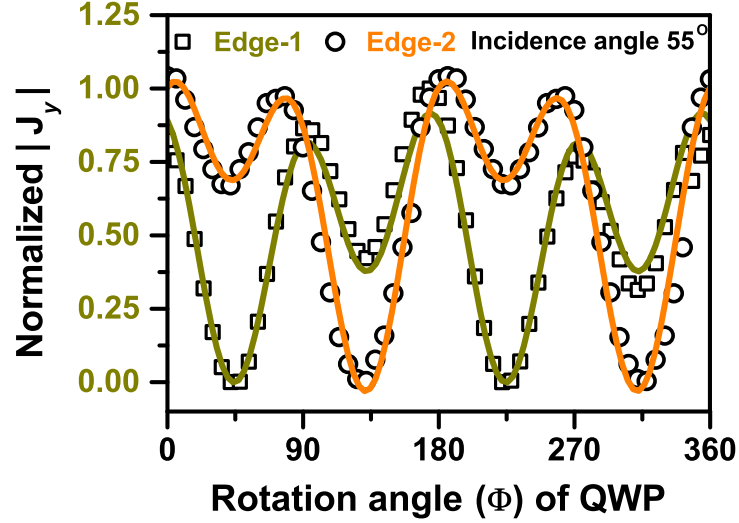}
	\caption{Comparison of the magnitudes of normalized net photocurrent $|J_y|$ at the two edges for $55\degree$ angle of incidence. The net photocurrent has opposite signs (flows in the opposite directions) at the two edges.}	\label{fig7}
\end{figure}

\begin{figure}[htb]
	\centering
	\includegraphics[clip,height=7.0cm]{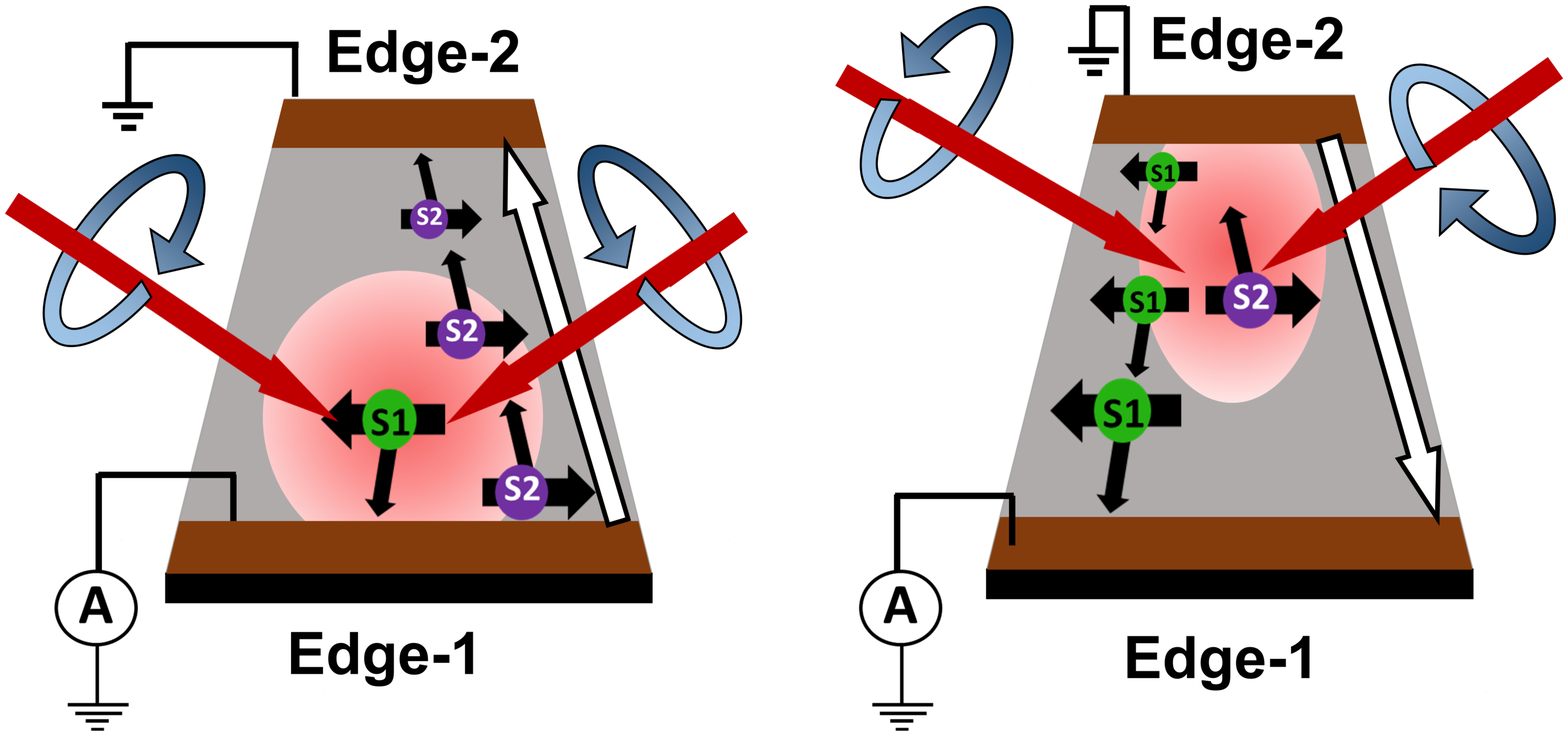}
	\caption{Photovoltage generation in presence of thermoelectric potential.}	\label{fig8}
\end{figure}

Let us now look at Fig.~\ref{fig6} and focus our attention on one of the two edges, say edge-1. There, it is clearly noticeable that $J_C$ and $J_1$ flip their signs as soon as we change the angle of incidence from positive (say $+55\degree$) to negative (say $-55\degree$) or vice versa. Interestingly,  $J_2$ and $D$ are unresponsive to the direction of the angle of incidence and does not change sign. The same is observed for edge-2 as well. This suggests that  $J_C$ and $J_1$ may have the same origin and it comes from Dirac like surface states. On the other hand, $J_2$ and $D$ will have a thermal origin, mainly brought about by the bulk carriers. For further confirmation let us compare the two edges, where it is evident that $J_2$ and $D$ reverse their signs as we go from one edge to the other. This can be understood if we consider thermal gradient plays a crucial role here, as it is oppositely directed at the two edges. Though the thermal gradient (in the two cases) can not change the helicity of the surface component, it can enhance the photocurrent in one case in comparison to the other~\cite{Yan2014}. 

In Fig.~\ref{fig7} we have plotted the normalized photocurrent when the beam spot is focused near edge-1 and edge-2, respectively, with an angle of incidence of $55\degree$ of the incoming light beam. It is clearly evident from this plot that the magnitude of photocurrent excited by the RCP ($\Phi=45\degree$) photons is lower than that excited by the LCP ($\Phi=135\degree$) photons near edge-1. On the other hand,  near edge-2, the magnitude of photocurrent excited by the LCP ($\Phi=135\degree$) photons is lower than that excited by the RCP ($\Phi=45\degree$) excitation. Note that the net photocurrent has opposite sign (flows in opposite directions) at the two edges. We may try to explain these observations in reference to the schematics shown in Fig.~\ref{fig8}. In this diagram, when the laser spot is focused near edge-1, the thermoelectric potential is directed from edge-1 to edge-2, as indicated by the white arrow. Now for the RCP ($\Phi=45\degree$) photons with positive incident angle, say $+55\degree$, or the LCP ($\Phi=135\degree$) photons with angle of incidence of $-55\degree$, both will have the same in-plane angular momentum component which preferentially excite one spin here, which we are naming as S1 for edge-1. Due to the coupling with momenta and as a spin-imbalance is created, the unexcited spin S2 will move toward edge-2. Also the thermal gradient directed towards edge-2 will further accelerate the movement of spin S2, giving rise to spin filtered enhanced photocurrent. However, if we exchange the photon polarization there will be no enhancement. This is because the LCP ($\Phi=135\degree$) photons with positive angle of incidence, say $+55\degree$ or the RCP ($\Phi=45\degree$) photons with negative angle of incidence, say $-55\degree$ will excite S2 spin having momentum towards edge-2, which is directed opposite to the thermoelectric potential. Now holding this combination of photon polarization and angle of incidence, if we shift the beam spot near edge-2, again we will get enhanced photocurrent but in reverse direction with opposite spin in comparison to the previous case when the illumination was focused near edge-1. This is because thermoelectric potential is now pointing towards edge-1 matching with S1 spin momentum. This explains the observation noted in Fig.~\ref{fig7}. 

\section{Conclusion}
In this report, photocurrent measurements on PLD-grown polycrystalline Sb$_2$Te$_2$Se thin films show all the unique properties of topological surface states. Helical photocurrent has been detected by shining  circularly polarized light at the center of the sample, far away from the contact pads, to avoid any thermoelectric influence. Interestingly, we observe that the helicity of the photocurrent can be tuned not only by changing the polarization of the excitation beam but also with the angle of incidence  of illumination. Subsequently, the effect of thermoelectric potential on spin transport has been examined by shifting laser illumination towards the two edges of the sample and by reversing the angle of incidence.  Systematic study showed enhanced helical photo-thermoelectric response as well as sign reversal in the two cases, when the beam spot is at the two opposite edges as thermal gradient plays a vital role here.

\acknowledgments{Authors acknowledge Ministry of Education, Govt. of India.
	S. M. acknowledges Council of Scientific \& Industrial Research (CSIR), India for
	research fellowship (09/921(0166)/2017-EMR-I). S. R. acknowledges Department of Science and Technology (DST), India for INSPIRE fellowship.}

\appendix

\section{X-ray diffraction study}
\begin{figure}[htb]
	\includegraphics[clip,width=0.6\linewidth]{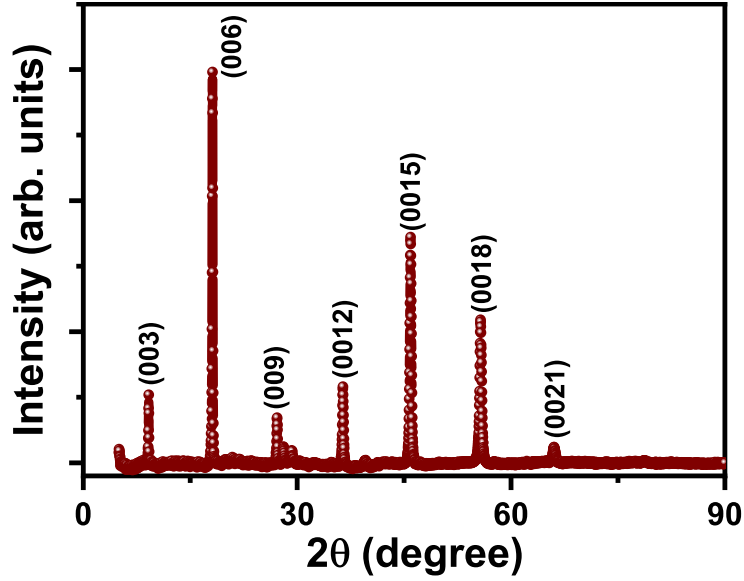}
	\caption{XRD spectrum of the sample}\label{XRD}       
\end{figure}

Figure~\ref{XRD} shows the x-ray diffraction (XRD) pattern of the studied STS thin film. The Miller indices ($h\,k\,l$) are shown in the figure and only the ($0\,0\,l$) peaks can be seen prominently. This enunciate a highly c-axis oriented granular layered structure of the studied sample. The peak position analysis coupled with the corresponding ($0\,0\,l$) values, yields an average lattice constant (c) of   29.47\AA, which agrees well with the earlier reports~\cite{Anderson1974}. We have used the Bragg diffraction formula for estimating the value of $d$ (spacing between successive crystal planes), given by, $n\lambda = 2d_{hkl} \sin\theta_{hkl}$. 
Here, $n=1$ and the wavelength of the x-ray used was  $\lambda=1.5406$\AA. The subscripts $hkl$ indicate the corresponding peaks that have been indexed successfully. Subsequently, since $h=k=0$ for these c-axis oriented grains, we used the formula, $d_{hkl} = c/l$ to calculate $c$. 
We calculated the values of $c$ for a few peaks and then took the average. 

\section{Transport measurements}
\begin{figure}[htb]
	\includegraphics[clip,width=\linewidth]{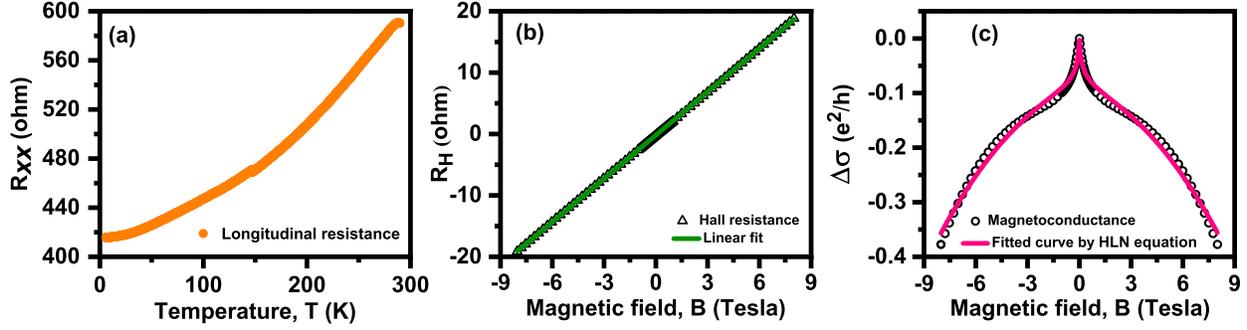}
	\caption{(a) Longitudinal resistance increases with temperature in the range of 5-295~K, showing metallic behavior of the sample. (b) Hall resistance measured in the magnetic field ($B$) range of $\pm9$~Tesla shows a linear dependence on $B$. (c) Magnetoconductance measurement in the magnetic field ($B$) range of $\pm9$~Tesla, showing cusp-like behavior near $B=0$.} \label{transport}       
\end{figure}

Transport measurements were performed on this sample in a cryogen-free magnet from Cryogenic Ltd., UK, to investigate the electronic properties of as deposited thin films. The graph of the temperature dependence of resistance, shown in Fig.~\ref{transport}~(a), clearly reveals the metallic nature of our sample. Temperature was varied in the range of 5-295~K. Subsequently, Hall measurements for these films have been carried out at 7~K. Linear dependency and a positive slope of Hall resistance with magnetic field as shown in Fig.~\ref{transport}~(b), captured the p-type signature of the bulk carriers (since the behavior is metallic). This shows that the Fermi level is lying in the valence band. From the slope, we have calculated the bulk carrier density which is around $3\times 10^{18}$~cm$^{-3}$. To confirm the existence of the surface states, we have measured the magnetoconductance (MC) at 7~K. Cusp-like  negative MC around zero magnetic field clearly observable in Fig.~\ref{transport}~(c), represents the weak anti-localization behavior which points to the existence of topological surface states. The MC data has been fitted to the Hikami-Larkin-Nagaoka (HLN) equation~\cite{Hikami1980} for TI:
\begin{equation}\label{HLN}
	\Delta\sigma(B) = -\alpha   \frac{e^2}{2\pi^2\hbar} \left[\Psi\left(\frac{1}{2}+\frac{\hbar}{4l_{\phi}^{2}eB}\right)-\mathrm{ln}\left(\frac{\hbar}{4l_{\phi}^{2}eB}\right)\right]
\end{equation}
along with an additional classical $B^2$ term due to the presence of bulk carriers. Here, $\sigma$ is the sample conductivity, $l_{\phi}$ is the phase coherence length, $B$ is magnetic field applied perpendicular to the sample surface, $\alpha$ is the number of independent conduction channels and $\Psi(x)$  represents the Digamma function that takes care of the singularity at the zero field due to the logarithmic function. Our experimental results fit well with this theoretical formula, suggesting presence of the topological surface states. 

%
%

\section{Functional dependence of $J_2$ on the angle of incidence $\theta$}
We have fitted the photocurrent component $J_2$ by using the formula derived following the treatment given in Ref.~\onlinecite{Plank2016}. The actual formula by considering the linear photo galvanic effect (LPGE) and photon drag effect (PDE) measured along the direction transverse to the plane of incidence (y-direction in our experimental configuration) for a linear polarization dependent photocurrent was  
\begin{equation}\label{actual_J_2_fitting}
	J_y = J_{\text{off}}-\cos2\alpha \frac{E_0^2}{2}\left[\left(\chi-T_\perp q \cos\theta\right)\left(t_s^2 + t_p^2 \cos^2\theta\right) + T_\parallel qt_p^2 \sin^2\theta \cos\theta\right]
\end{equation}
where $J_{\text{off}}$ was a polarization independent offset, $\chi$ is the linear photogalvanic term, $t_s$ and $t_p$ are the Fresnel transmission coefficients for s- and p-polarized light, $q$ is the magnitude of photon wave vector, $T_\perp$ and $T_\parallel$ are the photon drag coefficients for normal and oblique incidence, $\theta$ is the angle of incidence of the excitation light beam, and $\alpha$ is the rotation angle of the half wave plate, modulating the linear polarization of the illumination. In our case $\chi$ is zero as LPGE does not arise the in inversion-symmetric bulk. Also, the polarization independent offset term $J_{\text{off}}$ can be absorbed in the thermal contribution $D$. So, the reduced equation by considering PDE only is
\begin{equation}\label{J_2_fitting}
	J_{2}(\theta) = \cos4\Phi \left[A\cos\theta \left(B+\cos^2\theta\right) + C\sin^2\theta\cos\theta \right], 
\end{equation}
where, $A = E_0^2 T_\perp q t_p^2/2$, $B = t_s^2/t_p^2$, $C = E_0^2 T_\parallel q t_p^2/2$, and $\Phi$ is the rotation angle of the quarter wave plate used in our experiment to modulate the incoming light polarization. Here, the $\cos2\alpha$ term in Eq.~\eqref{actual_J_2_fitting} for the rotating half wave plate has been changed to $\cos4\Phi$ in Eq.~\eqref{J_2_fitting} for the rotating quarter wave plate, as it modulates the linear polarization as $\cos4\Phi$. 

Though, the photon drag is allowed for both the surface and bulk states, here we are claiming that it only comes from the bulk. The reason behind this is that the actual derivation~\cite{Plank2016} had considered the wrapping effect which is $C_{3v}$ symmetric. But in the low energy limit, the surface Dirac cone is rotationally symmetric around the $\Gamma$ point in $k$-space. This can be verified from our observation, where the helicity-dependent component $J_c$ of the photocurrent is zero for perpendicular incidence of light. This suggests that there is no out of plane spin component that stems from the hexagonal warping effect. This excludes any surface contribution to photon drag.

\section{Direction of helicity-dependent photocurrent}
\begin{figure}[htb]
	\includegraphics[clip,width=0.95\linewidth]{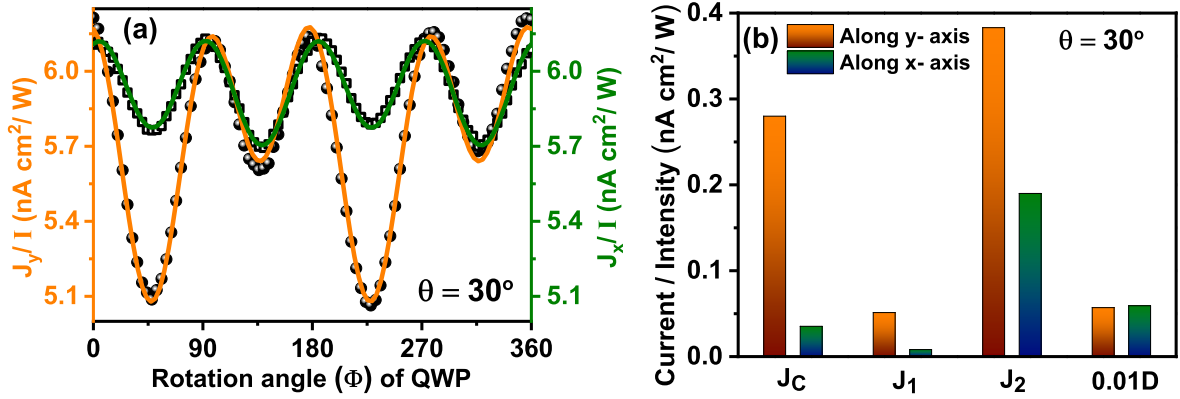}
	\caption{(a) Variation of photocurrent measured parallel (along x-axis) and perpendicular (along y-axis) to the plane of incidence of the excitation laser beam. (b) The bar graph compares various components of the net photocurrent measured along x- and y-axes. The large component $D$ is scaled down to $0.01D$ for clarity in the plot.}\label{xy photocurrent}       
\end{figure}

Helicity-dependent photocurrent is generated in the direction transverse to the plane of incidence of the excitation light beam due to the spin-orbit coupling of the carriers in the surface states. Here, we have measured this photocurrent in both transverse (along y-axis) as well as  longitudinal (along x-axis) directions, while changing the polarization state of the excitation laser beam by a rotating QWP. The excitation beam was incident at the central part of the sample at an angle of $30\degree$ with respect to the surface normal. Variation of the net photocurrent with rotation angle of the QWP is shown in Fig.~\ref{xy photocurrent}~(a).  After fitting the data with Eq.~\eqref{photocurrent} described in the main article, we obtained the values of various components of the measured photocurrent. These are plotted in Fig.~\ref{xy photocurrent}~(b), where only the absolute values are reported for a better understanding and comparison. From Fig.~\ref{xy photocurrent}~(b), it is obvious that the value of $J_c$ along y-direction is an order higher than that along the x-direction. This tells us that the helicity-dependent component ($J_c$) is predominantly operating along the transverse y-direction.

\bibliography{Helicity-dependent_photocurrent_topological_insulator}
\end{document}